\documentstyle[aps,prd,floats]{revtex}

\newlength{\cw}
\def\nad#1{\mbox{\smash{\oalign{$#1$\crcr\hidewidth$\mathchar"017E$
\hidewidth}}}}
\def\arraystretch{1.2}

\begin{document}

\title{\marginpar{\vspace{-.5in}\hspace{-1in}\small KFT U{\L} 2/92}
Quantization of the  tachyonic field}
\author{Jakub Rembieli\'nski\thanks{This work is supported under the
{\L}\'od\'z University grant no.\ 457.}}
\address{Katedra Fizyki Teoretycznej, Uniwersytet {\L}\'odzki\\
ul.~Pomorska 149/153, 90--236 {\L}\'od\'z, Poland\/%
\thanks{{\it E-mail address\/}: jaremb@mvii.uni.lodz.pl}}
\date{October 11, 1994}
\maketitle

\begin{abstract}
A consistent quantization scheme for imaginary-mass field is proposed.  It is
related to an appriopriate choice of the synchronization procedure
(definition of time), which guarantee an absolute causality.  In that
formulation a possible existence of field exctitations (tachyons) distinguish
an inertial frame (tachyon privileged frame of reference) {\em via\/}
spontaneous breaking of the so called synchronization group.
\end{abstract}

\section{Introduction}
Some attention in the literature over last decades, related to the question
of existence of faster-than-light particles, has been lacking in view of the
apparent conflict with the causality principle.  Irrespective of an attempt
to reconcile the notion of superluminal objects with causality on the
classical and/or semiclassical level \cite{Rec1,OR,Rec2}, it is commonly
believed that there is no respectable tachyonic quantum field theory at
present (for some efforts in this direction see \cite{KK,KK2}).

However, in the last time we observe a return of interest in tachyons.  This
is related to some recent experimental data \cite{tab,Ass} indicating that
the square of the muon neutrino mass seems to be negative\footnote{In the
paper by Chodos {\em et al.\/} \cite{CHK} it has been suggested that the muon
neutrino might possible be a fermionic tachyon. (See also \cite{Bie}).}.
Experimental data for the electron neutrino prefer also a negative value
for its mass square, but there are not so transparent \cite{tab}.

On the other hand the admitance of space-like four-momentum eigenstates can
possibly improve situation in quantum field theory by the weakness of the
spectral condition.  Furthermore non-localizability of tachyonic modes
\cite{Bar} may moderate QFT divergences.  It is also noticeable that a
tachyonic condensate is an immanent point of superstring models
\cite{KS,KP,CG,DS,Tse,MM}.

This paper is motivated by the problems mentioned above.   Here we make the
first step in this direction by a consistent quantization of a scalar
imaginary-mass field.  Our quantization scheme is related to a nonstandard
procedure of synchronization of clocks proposed in \cite{Tan,Cha1,Cha2,Rem1}.
This procedure allows us to introduce the notion of a coordinate time
appriopriate to the definition of the universal notion of causality in
agreement with special relativity. The main results can be summarized as
follows:
\begin{itemize}
\item
The relativity principle is formulated in the framework of a nonstandard
synchronization scheme (the Chang--Thangherlini (CT) scheme).  The absolute
causality holds for all kinds of events (time-like, light-like, space-like).
\item
For bradyons and luxons our scheme is fully equivalent to the standard
formulation of special relativity.
\item
For tachyons it is possible to formulate covariantly  proper initial
conditions.
\item
There exists a (covariant) lower bound of energy for tachyons.
\item
The paradox of ``transcendental'' tachyons is solved.
\item
Tachyonic field can be consistently quantized using the CT synchronization
scheme.
\item
The familiar ``reinterpretation principle'' \cite{Fei} cannot be unitarily
implemented on the quantum level; moreover there is no such necessity.
\item
Tachyons distinguish a preferred frame {\em via\/} mechanism of the
spontaneous symmetry breaking \cite{Rem2}.
\end{itemize}

\section{Preliminaries}\label{pre}
As is well known, in the standard framework of the special relativity,
space-like geodesics do not have their physical counterparts.  This is an
immediate  consequence of the assumed causality principle which admits
time-like and light-like trajectories only.

In the papers by Terletsky \cite{Ter}, Tanaka \cite{Tnk},
Sudarshan {\em et al.\/} \cite{BDS}, Recami {\em et al.\/}
\cite{Rec1,OR,Rec2} and Feinberg \cite{Fei} the causality problem has been
reexamined and a physical interpretation of space-like trajectories was
introduced.  However, every proposed solution raises new unanswered questions
of the physical or mathematical nature \cite{T&M}.  The difficulties are
specially frustrating on the quantum level \cite{KK,KK2,Nak}.  It is rather
evident that a consistent description of tachyons lies in a proper extension
of the causality principle. Note that interpretation of the space-like world
lines as physically admissible tachyonic trajectories favour the
constant-time initial hyperplanes.  This follows from the fact that only
such surfaces intersect each world line with locally nonvanishing slope once
and only once.  Unfortunatelly, the instant-time hyperplane is {\em not a
Lorentz-covariant notion\/} which is just the source of many troubles with
causality.

The first step toward a solution of this problem can be found in the papers
by Chang \cite{Cha1,Cha2,Cha3}, who introduced four-dimensional version of
the Tangherlini transformations \cite{Tan}, termed the Generalized Galilean
Transformations (GGT).  In \cite{Rem1} it was shown that GGT, extended to
form a group, are hidden (nonlinear) form of the Lorentz group
transformations with $SO(3)$ as a stability subgroup.  Moreover, a difference
with the standard formalism lies in a nonstandard choice of the
synchronization procedure.  As a consequence a constant-time hyperplane is a
covariant notion.  In the following we will call this procedure of
synchronization the {\em Chang--Tangherlini synchronization scheme}.

It is important to stress the following two well known facts from special
relativity: (a)~the definition of a coordinate time depends on the
synchronization scheme \cite{Rei,VT,Var}, (b)~synchronization scheme is a
convention, because no experimental procedure exists which makes it possible
to determine the one-way velocity of light without use of superluminal
signals \cite{Jam}. Notice that a choice of a synchronization scheme {\em
does not affect the assumptions of special relativity but evidently it can
change the causality notion}, depending on the definition of the coordinate
time.

Following Einstein, intrasystemic synchronization of clocks in their
``setting'' (zero) requires a definitional or conventional stipulation (for
discussion see Jammer \cite{Jam} and Sj\"odin \cite{Sjo}).  Really, to
determine one-way light speed it is necessary to use synchronized clocks (at
rest) in their ``setting'' (zero)\footnote{Obviously it is possible to
synchronize clocks in their rate withouth knowledge of the one-way light
speed \cite{And}.}.
On the other hand to synchronize clocks we should know the one-way light
velocity.  Thus we have a logical loophole.  In other words no experimental
procedure exists (if we exclude superluminal signals) which makes possible to
determine unambigously and without any convention the one-way velocity of
light. Consequently, an {\em operational meaning has the average value of the
light velocity around closed paths only}.  This statement is known as the
conventionality thesis \cite{Jam}. Following Reichenbach \cite{Rei}, two
clocks $\sf A$ and $\sf B$ stationary in the points $A$ and $B$ of an
inertial frame are defined as being synchronous with help of light signals if
$t_B=t_A+\varepsilon_{AB}(t'_A-t_A)$.  Here $t_A$ is the emission time of
light signal at point $A$ as measured by clock $\sf A$, $t_B$ is the
reception-reflection time at point $B$ as measured by clock $\sf B$
and $t'_A$ is the reception time of this light signal at point $A$ as
measured by clock $\sf A$.  The so called synchronization coefficient
$\varepsilon_{AB}$ is an arbitrary number from the open interval $(0,1)$.
In principle it can vary from point to point.  The only conditions for
$\varepsilon_{AB}$ follow from the requirements of symmetry and transitivity
of the synchronization relation.  Note that
$\varepsilon_{AB}=1-\varepsilon_{BA}$. The one-way velocities of light
from $A$ to $B$ ($c_{AB}$) and from $B$ to $A$ ($c_{BA}$) are given by
\[
c_{AB}=\frac{c}{2\varepsilon_{AB}},\quad
c_{BA}=\frac{c}{2\varepsilon_{BA}}.
\]
Here $c$ is the round-trip average value of the light velocity.  In standard
synchronization $\varepsilon_{AB}=\frac{1}{2}$ and consequently $c=c_{AB}$
for each pair $A$, $B$.

The coventionality thesis states that from the operational point of view the
choice of a fixed set of the coefficients $\varepsilon$ is a convention.
However, the explicit form of the Lorentz transformations will be
$\varepsilon$-dependent in general.  The question arises:  Are equivalent
notions of causality connected with different synchronization schemes?
As we shall see throughout this work the answer is {\em negative\/} if we
admit tachyonic world lines.  In other words, the causality requirement,
logically independent of the requirement of the Lorentz covariance, can
contradict the conventionality thesis and  consequently it can prefer a
definite synchronization scheme, namely CT scheme.

\section{The Chang--Tangherlini synchronization}\label{CTsyn}
As was mentioned in Section \ref{pre}, in the paper by Tangherlini \cite{Tan}
a family of inertial frames in $1+1$ dimensional space of events was
introduced with the help of transformations which connect the time
coordinates by a simple (velocity dependent) rescaling.  This construction
was generalized to the $1+3$ dimensions by Chang \cite{Cha1,Cha2}. As was
shown in the paper \cite{Rem1}, the Chang--Tangherlini inertial frames can be
related by a group of transformations isomorphic to the orthochronous Lorentz
group.  Moreover, the coordinate transformations should be supplemented by
transformations of a vector-parameter interpreted as the velocity of a
privileged frame.  It was also shown that the above family of frames is
equivalent to the Einstein--Lorentz one; (in a contrast to the interpretation
in \cite{Cha1,Cha2}).  A difference lies in another synchronization procedure
for clocks \cite{Rem1}.

In this Section we derive a realization of the Lorentz group given in
\cite{Rem1} in a systematic way \cite{Rem2}.  Furthermore we will discuss
physical content of our formalism.

Let us start with a simple observation that the description of a family of
(relativistic) inertial frames in the Minkowski space-time is not so
convenient and natural.  Instead, it seems that the geometrical notion of
bundle of frames is more natural.  Base space is identified with the space of
velocities; each velocity marks out a coordinate frame.  Indeed, from the
point of view of an observer (in a fixed inertial frame) all inertial frames
are labelled by their velocities with respect to him.  Therefore, in
principle, to define the transformation rules between frames, we can use,
except of coordinates, also this vector-parameter, possibly related to
velocities of frames with respect to a (formally) distinguished observer.
Because we adopt Lorentz covariance, we can use a time-like fourvelocity
$u_E$; subscript $E$ means Einstein--Poincar\'e synchronization\footnote{In
the papers by Chang \cite{Cha1,Cha2,Cha3} it was used some kinematical
objects with an unproper physical interpretation \cite{SP,FN}.  For this
reason we should be precise in the nomenclature related to different
synchronizations.} (EP synchronization) i.e.\ we adopt, at this moment, the
standard transformation law for $u_E$
\[
u_E'=\Lambda u_E
\]
where $\Lambda$ is an element of the Lorentz group $L$.

Below we list our main assumptions:
\begin{enumerate}
\item
Coordinate frames are related by a set of transformations isomorphic to the
Lorentz group ({\bf Lorentz covariance}).
\item
The average value of the light speed over closed paths is constant $(c)$ for
all inertial observers ({\bf constancy of the round-trip light velocity}).
\item
With respect to the rotations $x^0$ and $\vec{x}$ transform as $SO(3)$
singlet and triplet respectively ({\bf isotropy}).
\item
Transformations are linear with respect to the coordinates ({\bf affinity}).
\item
(*) We admit an additional set of parameters $u_E$ ({\bf the base
space} for a bundle of inertial frames).
\item
(*) Instant-time hyperplanes are covariant under coordinate transformations
({\bf absolute causality}).
\end{enumerate}
We see that assumptions labelled by star * are essentialy new.

\subsection{Derivation of the Lorentz group transformation rules in the CT
synchronization}\label{Lor}
According to our assumptions, transformations between two coordinate frames
$x^{\mu}$ and ${x'}^\mu$ have the following form
\begin{equation}
x'(u_E')=D(\Lambda,u_E)x(u_E).                                     \label{A1}
\end{equation}
Here $D(\Lambda,u_E)$ is a real (invertible) $4\times4$ matrix, $\Lambda$
belongs to the Lorentz group and $u_{E}^{\mu}$ is assumed to be a Lorentz
four-vector, i.e.\
\begin{equation}
u_E'=\Lambda u_E,\quad {u_E}^2=c^2>0.                              \label{A2}
\end{equation}
The physical meaning of $u_E^\mu$ will be explained later.  It is easy to
verify that the transformations (\ref{A1}--\ref{A2}) constitute a realization
of the Lorentz group if the following composition law holds
\begin{equation}
D({\Lambda}_2,{\Lambda}_1u_E)D({\Lambda}_1,u_E)=
D({\Lambda}_2{\Lambda}_1,u_E).                                     \label{D2}
\end{equation}
Now {\em we demand that\/} $(x^\mu)\equiv(x^0,\vec{x})$ {\em transform under
subgroup of rotations as singlet + triplet\/} ({\em isotropy condition\/})
i.e.\ for $R\in SO(3)$
\begin{equation}
\Omega\equiv D(R,u_E)=\pmatrix{1&0\cr 0&R}.                        \label{A3}
\end{equation}
{}From eqs.\ (\ref{A1}--\ref{D2}) we see that the identity and the inverse
element have the form
\begin{mathletters}
\label{A4}
\begin{equation}
I=D(I,u_E),
\end{equation}
\begin{equation}
D^{-1}(\Lambda,u_E)=D(\Lambda^{-1},\Lambda u_E).
\end{equation}
\end{mathletters}
Using the familiar Wigner trick we obtain that
\begin{equation}
D(\Lambda,u_E)=T(\Lambda u_E)\Lambda T^{-1}(u_E),                  \label{A5}
\end{equation}
where the real matrix $T(u_E)$ is given by
\begin{equation}
T(u_E)=D(L_{u_E},\tilde u_E)L_{u_E}^{-1}.                          \label{A6}
\end{equation}
Here $\tilde u_E=(c,0,0,0)$ and $L_{u_E}$ is the boost matrix:
$u_E=L_{u_E}\tilde u_E$.  We use the following parametrization of the matrix
$L_{u_E}$
\[
L_{u_E}=\left(\begin{array}{c|c}
\displaystyle\frac{u^0_E}{c}
&\displaystyle\frac{\vec u_E^{\rm T}}{c}\\[1ex]
\hline
\displaystyle\frac{\vec u_E}{c}
&\displaystyle I+\frac{\vec u_E\otimes\vec u_E^{\rm T}}{c^2\left(1+
\displaystyle\frac{u^0_E}{c}\right)}
\end{array}
\right).
\]

Note that the transformations (\ref{A1}--\ref{A2}) leave the bilinear form
$x^{\rm T}(u_E)g(u_E)x(u_E)$, where the symmetric tensor $g(u_E)$ reads
\begin{equation}
g(u_E)=(T(u_E)\eta T^{\rm T}(u_E))^{-1},                           \label{A7}
\end{equation}
invariant.
Here $\eta$ is the Minkowski tensor and the superscript $^{\rm T}$ means
transposition.

Now we determine the matrix $T(u_E)$. To do this we note that under rotations
\[
T(\Omega u_E)=\Omega T(u_E)\Omega^{-1},
\]
so the most general form of $T(u_E)$ reads
\begin{equation}
T(u_E)=\left(\begin{array}{c|c}
a(u^0_E)&b(u^0_E)\vec u_E^{\rm T}\\[1ex]
\hline
d(u^0_E)\vec{u_E}& e(u^0_E)I+(\vec u_E\otimes\vec u_E^{\rm T}) f(u^0_E)
\end{array}\right),                                                \label{A8}
\end{equation}
where $a$, $b$, $d$, $e$ and $f$ are some functions of $u^0_E$.
Inserting eq.\ (\ref{A8}) into eq.\ (\ref{A7}) we can express the metric
tensor $g(u_E)$ by $a$, $b$, $d$, $e$ and $f$.  In a three dimensional flat
subspace we can use an orthogonal frame (i.e.\ $(g^{-1})^{ik}=-\delta^{ik};$
$i,k=1,2,3$), so we obtain
\begin{equation}
e(u^0_E)=1,\quad d^2=(2-f\vec u_E^2)f.                             \label{A9}
\end{equation}
Furthermore, from the equation of null geodesics, $dx^{\rm T}g\,dx=0$, we
deduce that the light velocity $\vec c$ in the direction $\vec n$
($\vec n^2=1$) is of the form
\begin{equation}
\vec c=c\vec n\left(\sqrt{\alpha+\beta^2\vec u_E^2}
-\beta\vec u_E\vec n\right)^{-1},                                 \label{A10}
\end{equation}
where $\alpha=a^2-b^2\vec u_E^2$, $\beta=ad-b(1+f\vec u_E^2)$. From eq.\
(\ref{A10}) we see that the synchronization convention depends on the
functions $\alpha$ and $\beta$ only.  Now, because $a$, $b$ and $d$ can be
expressed as functions of $\alpha$, $\beta$ and $f$ and we are interested in
{\em esentially different synchronizations\/} only, we can choose
\begin{equation}
f=0,                                                             \label{A11a}
\end{equation}
so
\begin{equation}
d=0,\quad\beta=-b,\quad\alpha=a^2-b^2\vec u_E^2.                 \label{A11b}
\end{equation}
Finally, from (\ref{A10}--\ref{A11b}) the average value of $|\vec c|$ over a
closed path is equal to
\[
\langle|\vec c|\rangle_{\text{cl.\ path}}=\frac{c}{a}.
\]
Because {\em we demand that the round-trip light velocity\/}
($\langle|\vec c|\rangle_{\text{cl.\ path}}=c$) {\em be constant}, we obtain
\begin{equation}
a=1.                                                              \label{A12}
\end{equation}
Summarizing, $T(u_E)$ has the form
\begin{equation}
T(u_E)=\left(\begin{array}{c|c}
1&b(u^0_E)\vec u_E^{\rm T}\\[1ex]
\hline
0&I\end{array}\right),                                            \label{A13}
\end{equation}
while the light velocity
\begin{equation}
\vec c=c\vec n\left(1+b\vec u_E\vec n\right)^{-1},                \label{A14}
\end{equation}
so the Reichenbach coefficient reads
\begin{equation}
\varepsilon(\vec n,\vec u_E)
=\frac{1}{2}\left(1+b\vec u_E\vec n\right).                       \label{A15}
\end{equation}
In special relativity the function $b(u^0_E)$ distinguishes between different
synchronizations.  Choosing $b(u^0_E)=0$ we obtain $\vec c=c\vec n$,
$\varepsilon=\frac{1}{2}$ and the standard trasformation rules for
coordinates: $x'=\Lambda x$.  On the  other hand, if we demand that the
instant-time hyperplane $x^0={\rm constant}$ be an invariant notion, i.e.\
that ${x'}^0={D(\Lambda, u_E)^0}_0 x^0$ so ${D(\Lambda, u_E)^0}_k=0$, then
from eqs.\ (\ref{A5}, \ref{A13}) we have
\begin{equation}
b(u^0_E)=-\frac{1}{u^0_E}.                                        \label{A16}
\end{equation}
In the following we restrict ourselves to the above case defined by eq.\
(\ref{A16}).  Notice that $\vec u_E/u^0_E$ can be expresed by a
three-velocity $\vec\sigma_E$
\begin{equation}
\frac{\vec u_E}{u^0_E}=\frac{\vec\sigma_E}{c}                     \label{A17}
\end{equation}
with $0\leq|\vec\sigma_E|<c$. Therefore
\begin{equation}
T(u_E)\equiv T(\vec\sigma_E)=\left(\begin{array}{c|c}
1&\displaystyle-\frac{\vec\sigma^{\rm T}_E}{c}\\[1ex]
\hline
0&I\end{array}\right).                                           \label{A18a}
\end{equation}
Thus we have determined the form of the transformation law (\ref{A1}) in this
case.  Now, according to our interpretation of the freedom in the Lorentz
group realization as the synchronization convention freedom, there should
exists a relationship between $x^\mu$ coordinates and the
Einstein--Poincar\'e ones.  In fact, the matrix $T$ relates both
synchronizations {\em via\/} the formula
\begin{equation}
x=T(\vec\sigma_E)x_E.                                              \label{D3}
\end{equation}
Explicitly:
\begin{mathletters}
\label{D4}
\begin{equation}
x^0=x^0_E-\frac{\vec\sigma_E}{c}\vec x_E,
\end{equation}
\begin{equation}
\vec x=\vec x_E.
\end{equation}
\end{mathletters}
It is easy to check that $x_E$ transforms according to the standard law i.e.\
\begin{equation}
x_E'=\Lambda x_E.                                                  \label{D5}
\end{equation}
Now, by means of eq.\ (\ref{D4}) we obtain analogous relations between
differentials
\begin{mathletters}
\label{D6}
\begin{equation}
dx^0=dx^0_E-\frac{\vec\sigma_E}{c}\,d\vec x_E,
\end{equation}
\begin{equation}
d\vec x=d\vec x_E,
\end{equation}
\end{mathletters}
and consequently interrelations between velocities in both synchronizations;
namely
\begin{equation}
\vec v=\frac{\vec v_E}{1-\displaystyle\frac{\vec v_E\vec\sigma_E}{c^2}},
\label{D7}
\end{equation}
\begin{equation}
\vec v_E=\frac{\vec v}{1+\displaystyle\frac{\vec\sigma\vec v}{c^2}
\gamma^{-2}_0}.
                                                                   \label{D8}
\end{equation}
Here $\vec\sigma$ is the $\vec\sigma_E$ velocity in the CT synchronization,
i.e.
\begin{equation}
\vec\sigma=\frac{\vec\sigma_E}{1-\left(\displaystyle\frac{\vec\sigma_E}{c}
\right)^2},
                                                                   \label{D9}
\end{equation}
while $\gamma_0$ is defined as
\begin{equation}
\gamma_0=\left[\frac{1}{2}\left(1+\sqrt{1+\left(\displaystyle\frac{2
\vec\sigma}{c}\right)^2}\right)\right]^{1/2}.                     \label{D10}
\end{equation}
In the following we use also the quantity $\gamma(\vec v)$ defined as follows
\begin{equation}
\gamma(\vec v)=
\left|\left(1+\frac{\vec\sigma\vec v}{c^2}\gamma^{-2}_0\right)^2
-\left(\frac{\vec v}{c}\right)^{2}\right|^{1/2}.                  \label{D11}
\end{equation}
Now, taking into account eqs.\ (\ref{A14}, \ref{A16}, \ref{A17}) we see that
the light velocity $\vec c$ in the direction of a unit vector $\vec n$ reads
\begin{equation}
\vec c=\frac{c\vec n}{1-\displaystyle\frac{\vec n\vec\sigma_E}{c}},
\
label{D12}
\end{equation}
i.e.\ in terms of $\vec\sigma$ (see eq.\ (\ref{D9}))
\begin{equation}
\vec c=\frac{c\vec n}{1-\displaystyle\frac{\vec n\vec\sigma}{c}
\gamma_0^{-2}},                                                   \label{D13}
\end{equation}
so
\begin{equation}
\varepsilon(\vec n,\vec\sigma)=\frac{1}{2}\left(1-\frac{\vec n\vec\sigma}{c}
\gamma_0^{-2}\right).                                             \label{D14}
\end{equation}
We call the synchronization scheme defined by the above choice of the
Reichenbach coefficients the {\em Chang--Tangherlini synchronization}.
In terms of $\vec\sigma$, the matrix $T$ reads
\begin{equation}
T(\vec\sigma)=\left(\begin{array}{c|c}
1&\displaystyle-\frac{\vec\sigma^{\rm T}}{c}\gamma_0^{-2}\\[1ex]
\hline
0&I\end{array}\right).                                              \label{X}
\end{equation}

Let us return to the transformation laws (\ref{A1}) and (\ref{A2}).  By means
of the formulas (\ref{A5}, \ref{D8}) and (\ref{X}) we can deduce the
following explicit form of the Lorentz group transformations
(\ref{A1}--\ref{A2}) in the CT synchronization \cite{Rem1,Rem2}
\begin{description}
\item[Boosts] \hfil
\begin{mathletters}
\label{D15}
\begin{equation}
{x'}^0=\gamma x^0,
\end{equation}
\begin{equation}
{\vec x}'=\vec x+\frac{\vec V}{c}\left[\frac{\vec V\vec x}{c
\left(\gamma+\sqrt{\gamma^2+\left(\displaystyle\frac{\vec V}{c}\right)^2}
\right)}
-\frac{\vec\sigma\vec x}{c\gamma_0}-x^0\right]\gamma^{-1},
\end{equation}
\begin{equation}
{\vec\sigma}'=\vec\sigma\gamma^{-1}+\vec V\gamma^{-2}
\left[\frac{\vec V\vec\sigma}{c^2\left(\gamma+\sqrt{\gamma^2+
\left(\displaystyle\frac{\vec V}{c}\right)^2}\right)}
-\left(\frac{\vec\sigma}{c}\right)^2\gamma_0^{-1}-1\right].
\end{equation}
\end{mathletters}
Here $\gamma=\gamma(\vec{V})$ has the form (\ref{D11}).
\item[Rotations] (compare with (\ref{A3}))
\begin{mathletters}
\label{D16}
\begin{equation}
{x'}^0=x^0,
\end{equation}
\begin{equation}
{\vec x}'=R\vec x,
\end{equation}
\begin{equation}
{\vec\sigma}'=R\vec\sigma.
\end{equation}
\end{mathletters}
\end{description}
It is easy to see that a vector $\vec{V}$ appearing in the
transformations rules (\ref{D15}) is the relative velocity of the frame
$x'$ with respect to $x$, measured in the CT synchronization. Moreover,
from (\ref{D15}--\ref{D16}) we can deduce the meaning of the
vector-parameter $\vec{\sigma}$; namely $\vec{\sigma}$ is a velocity
of a fixed (formally privileged) frame as measured by an observer
which uses the coordinates $x(\vec\sigma)$.

Notice that the matrix $D$ (eq.\ (\ref{A1})) for Lorentz boosts reads
\begin{equation}
D(\vec{V},\vec{\sigma})=\left(\begin{array}{c|c}
\gamma & 0\\[1ex]
\hline
\displaystyle-\frac{\vec V}{c}\gamma^{-1} & I+
\displaystyle\frac{\vec V\otimes\vec V^{\rm T}}
{c^2\gamma\left[\gamma+\sqrt{\gamma^2+\displaystyle\frac{\vec V}{c}^2}
\right]}-\frac{\vec V\otimes\vec\sigma^{\rm T}}{c^2\gamma\gamma_0}
\end{array}\right).                                              \label{D20}
\end{equation}
For completeness, we give also the explicit form of the metric tensors
$g_{\mu\nu}(\vec{\sigma})$ and $g^{\mu\nu}(\vec{\sigma})$:
\begin{mathletters}
\begin{equation}
\left[g_{\mu\nu}(\vec\sigma)\right]=
\left(\begin{array}{c|c}
1 & \displaystyle\frac{\vec\sigma^{\rm T}}{c}\gamma_0^{-2}\\[1ex]
\hline
\displaystyle\frac{\vec\sigma}{c}\gamma_0^{-2}&
\displaystyle-I+\frac{\vec\sigma\otimes\vec\sigma^{\rm T}}{c^{2}}
\gamma_0^{-4}\end{array}\right),                                  \label{D17}
\end{equation}
\begin{equation}
\left[g^{\mu\nu}(\vec\sigma)\right]=
\left(\begin{array}{c|c}
\gamma_0^{-2}&\displaystyle\frac{\vec\sigma^{\rm T}}{c}\gamma_0^{-2}\\[1ex]
\hline
\displaystyle\frac{\vec\sigma}{c}\gamma_0^{-2}&
-I\end{array}\right).                                             \label{D18}
\end{equation}
\end{mathletters}
{}From (\ref{D18}) it is evident that the three-space is the Euclidean one.
Furthermore, the subset of transformations (\ref{D15}) defined by the
condition $\vec\sigma=0$ coincides exactly with the family of the
Chang--Tangherlini inertial frames \cite{Cha1,Cha2}.

\subsection{Causality and kinematics in the CT synchronization}\label{caus}
In this subsection we discuss shortly differences and similarities of
kinematical descriptions in both CT and EP synchronizations.  Recall that in
CT scheme {\em causality has an absolute meaning}.  This follows from the
transformation law (\ref{D15}) for the coordinate time: $x^0$ is rescaled by
a positive, velocity dependent factor $\gamma$.  Thus this formalism extends
the EP causality by allowing faster than light propagation.  It can be made
transparent if we consider the relation derived from  eq.\ (\ref{D6})
\begin{equation}
\frac{dx^0}{dx^0_E}=1-\frac{\vec\sigma_E\vec v_E}{c^2}.           \label{D19}
\end{equation}
For $|\vec v_E|\leq c$ we have $\frac{dx^0}{dx^0_E}>0$, whereas for
$|\vec v_E|>c$, $\frac{dx^0}{dx^0_E}$ can be evidently negative, which is a
consequence of an inadequacy of the EP synchronization in this situation.
Notice that subluminal (superluminal) signals in the EP synchronization
remain subluminal (superluminal) in the CT one too; indeed, as we see from
eqs.\ (\ref{D7}, \ref{D8}, \ref{D12}, \ref{D13}) the rate of the
corresponding velocities in the same direction $\vec{n}$ reads
\[
\frac{|\vec{v}|}{|\vec{c}|}=\left|\frac{\displaystyle\frac{v_E}{c}-
\frac{\vec{v}_{E}\vec{\sigma}_{E}}{c^2}}
{1-\displaystyle\frac{\vec{v}_{E}\vec{\sigma}_{E}}{c^2}}\right|<1\quad
\text{iff}\quad\left|\frac{v_E}{c}\right|<1.
\]

Let us consider in detail a space-like four-momentum $k^{\mu}$ transforming
under $D$ determined by eqs.\ (\ref{D15}--\ref{D16}) (see also (\ref{D20})).
Only in this case we can expect a real deviation from the standard
description.  Now, our $k$ satisfy
\begin{equation}
k^{2}=g_{\mu\nu}(\vec{\sigma})k^{\mu}k^{\nu}=
g^{\mu\nu}(\vec{\sigma})k_{\mu}k_{\nu}=-\kappa^2<0.               \label{D22}
\end{equation}
Because velocity of a particle has direct physical meaning we solve the
tachyonic dispersion equation (\ref{D22}) by means of the evident relations
\[
k^{\mu}=\frac{\kappa}{c}w^{\mu}
\]
with $w^{2}=-c^{2}$, and
\[
\frac{\vec{v}}{c}=\frac{d\vec{x}}{dx^0}=\frac{\vec{w}}{w^0}.
\]
Consequently the solution of eq.\ (\ref{D22}) reads
\begin{mathletters}
\label{D23}
\begin{equation}
k^{0}_{\pm}=\pm\kappa\gamma^{-1},
\end{equation}
\begin{equation}
\vec{k}_{\pm}=\pm\kappa\gamma^{-1}\frac{\vec{v}}{c},
\end{equation}
\end{mathletters}
where $\gamma=\gamma(\vec{v})$ is given by eq.\ (\ref{D11}).

Now, by means of (\ref{D17}) the covariant fourmomentum $k_{\mu}$ has the
form
\begin{mathletters}
\label{D24}
\begin{equation}
k_{0\pm}=\pm\kappa\gamma^{-1}\left(1+\gamma^{-2}_{0}
\frac{\vec{\sigma}\vec{v}}{c^2}\right),
\end{equation}
\begin{equation}
\nad{k}_{\pm}=\pm \kappa\gamma^{-1}\left[-\frac{\vec{v}}{c}+
\frac{\vec{\sigma}}{c}\gamma_{0}^{-2}\left(
1+\gamma^{-2}_{0}\frac{\vec{\sigma}\vec{v}}{c^2}\right)\right].
\end{equation}
\end{mathletters}
Recall that the generators of space-time translations are covariant, so
energy must be identified with $k_0$. To make a proper identification of
energy ($k_{0+}$ or $k_{0-}$), let us analyse the above formulas with the
help of  convenient parameters $\xi$, $s$ and $\varepsilon$
\begin{mathletters}
\label{D25}
\begin{equation}
\xi=\frac{|\vec{v}|}{|\vec{c}|} \in(1,\infty),\quad\text{(for tachyons)},
\end{equation}
\begin{equation}
s=\frac{|\vec{c}|}{c}\in(\frac{1}{2},\infty),
\end{equation}
\begin{equation}
\varepsilon=\gamma_{0}^{-2}\frac{\sigma}{c}\in\langle0,1).
\end{equation}
\end{mathletters}
Here $\vec{c}$ propagates in the direction of $\vec{v}$, i.e.\ $\vec{n}$
in eq.\ (\ref{D13}) is assumed to have the form $\vec{n}=\vec{v}/|\vec{v}|$.
In terms of $\xi$ and $s$
\begin{equation}
k_{0\pm}=\pm\kappa\frac{1+(s-1)\xi}{\sqrt{(\xi-1)[(2s-1)\xi+1]}}.
                                                                 \label{D26}
\end{equation}
We see that a proper choice for tachyon energy is $k_{0+}$; indeed $k_{0+}$
has a lower bound.  Moreover, this property is covariant because
$k^{0}_{+}$ is positive in that case and
$\varepsilon(k^{0}_{+})=1=\text{invariant}$.  Notice also that the lowest
value of energy $k_{0\text{min}}=\kappa(s-1)/(2s-1)^{1/2}$ depends only on
the light propagation characteristics in a given frame.  Thus in the CT
synchronization, contrary to the EP one, tachyonic energy is bounded from
below.  This fact is especially important because implies stability on the
quantum level.  Furthermore, invariance of the sign of $k^0$ allows the
covariant decomposition of the tachyon field on the creation and anihilation
part, so the Fock procedure can be applied.  For completeness, we give
also $k^0, \vec{k}$ and $\nad{k}$ in terms of $\xi$, $s$, $\varepsilon$
\begin{mathletters}
\label{QQ}
\begin{equation}
k^{0}=\frac{\kappa}{\sqrt{(\xi-1)[(2s-1)\xi+1]}},
\end{equation}
\begin{equation}
\vec{k}=\frac{\vec{n}\kappa\xi s}{\sqrt{(\xi-1)[(2s-1)\xi+1]}},
\end{equation}
\begin{equation}
\nad{k}=\frac{-\vec{n}\kappa\xi s+\vec{e}\kappa\varepsilon\left[
1+\xi(s-1)\right]}{\sqrt{(\xi-1)[(2s-1)\xi+1]}}.
\end{equation}
\end{mathletters}
Here $\vec{e}=\frac{\vec{\sigma}}{\sigma}$. Notice that
\begin{equation}
\left|\nad{k}\right|>\lim_{\xi\rightarrow\infty}\left|\nad{k}\right|=
\kappa\sqrt{\frac{1+(\varepsilon^2-2)\varepsilon^2\cos^2{\theta}}
{1-\varepsilon^2\cos^2{\theta}}},\label{QQQ}
\end{equation}
where $\cos{\theta}=\vec{n}\vec{e}$.

Finally, let us reexamine the problem of the so called ``transcendental''
tachyon.  To do this, recall the transformation law for velocities in the EP
synchronization \cite{And}
\begin{equation}
{\vec v}'_E=\frac{\gamma_{E}\vec{v}_E+\vec{V}_E\left[\displaystyle
\frac{\vec{V}_E\vec{v}_E}{c^2}\left(\gamma_E+1\right)^{-1}-1\right]}
{\displaystyle 1-\frac{\vec{V}_E\vec{v}_E}{c^2}},   \label{XX}
\end{equation}
where $\gamma_E=\sqrt{1-\left(\frac{\vec{V}_E}{c}\right)^2}$.

We observe that the denominator of the above transformation rule can vanish
for $\left|\vec{v}_E\right|>c$; Thus a tachyon moving with
$c<\left|\vec{v}_E\right|<\infty$ can be converted by a finite Lorentz map
into a ``transcendental'' tachyon with $\left|\vec{v}_E'\right|=\infty$.
This discontinuity is an apparent inconsistency of this transformation law;
namely {\em tachyonic velocity space does not constitute a representation
space for the Lorentz group\/}!

On the other hand, in the CT scheme, the corresponding transformation rule
for velocities follows directly from eq.\ (\ref{D15}) and reads
\begin{equation}
{\vec v}'=\gamma^{-1}\vec{v}+\gamma^{-2}\vec{V}\left[\displaystyle
\frac{\vec{V}\vec{v}}{c^2\left(\gamma+\sqrt{\gamma^2+\displaystyle
\frac{\vec{V}^2}{c^2}}
\right)}-\frac{\vec{\sigma}\vec{v}}{c^2}\gamma_{0}^{-2}-1\right],
\label{XXX}
\end{equation}
where $\gamma=\gamma(\vec{V})$. Thus contrary to eq.\ (\ref{XX}),
the transformation law (\ref{XXX}) is continous, does not ``produce''
``transcendental'' tachyons and completed by rotations, forms
(together with the mapping $\vec{\sigma}\rightarrow{\vec\sigma}'$) a
realization of the Lorentz group.

We finish this section with a Table \ref{tI} summarizing our results.

\begin{table}
\renewcommand{\arraystretch}{1}
\caption{Comparison of the special relativity descriptions of kinematics
in the Einstein--Poincar\'e and Chang--Tangherlini synchronization
schemes.\label{tI}}
\begin{center}
\squeezetable
\begin{tabular}{clcc}
\multicolumn{2}{l}{Synchronization scheme $\longrightarrow$}&
\multicolumn{1}{c}{Einstein--Poincar\'e}&
\multicolumn{1}{c}{Chang--Tangherlini}\\
\multicolumn{2}{l}{Description of $\downarrow$}&&\\
\hline\hline
\multicolumn{1}{c}{\parbox{.7\cw}{\centering BRADYONS \\ $k^2=\kappa^2$ \\
and LUXONS \\ $k^2=0$}}&&
\parbox{1.2\cw}{\raggedright Consistent causal kinematics, fully equvalent to
the CT description}&
\parbox{1.2\cw}{\raggedright Consistent causal kinematics, fully equvalent to
the EP description}\\
\hline
&uniwersal notion of causality&no&yes\\
&&&($\varepsilon(dx^0)={\rm inv}$)\\
\cline{2-4}&
covariant initial conditions&no&yes\\
&&&($x^0={\rm const}$)\\
\cline{2-4}
TACHYONS&invariant sign of $k^0$&no&yes\\
$k^2=-\kappa^2$&&&($\varepsilon(k^0)={\rm inv}$)\\
\cline{2-4}&
covariant lower bound of energy&no&yes\\
&&($k_0\to-\infty$)&($k_{0\text{min}}>-\infty$)\\
\cline{2-4}&
paradox of ``transcendental'' tachyons&inconsistency (discontinuity)&
consistent (continuous)\\
&&&picture
\end{tabular}
\end{center}
\end{table}

\subsection{Synchronization group and the relativity principle}\label{syn}
{}From the foregoing discussion we see that the CT synchronization prefers
a privileged frame corresponding to the value $\vec{\sigma}=0$ ({\em
relativistic ether\/}).
It is clear that if we forget about tachyons such a preference is only
formal; namely we can choose each inertial frame as a preferred one.

Let us consider two CT synchronization schemes, say $A$ and $B$,
under two different choices of privileged inertial frames, say
$\Sigma_A$ and $\Sigma_B$. Now, in each inertial frame $\Sigma$ two
coordinate charts $x_A$ and $x_B$ can be introduced, according to
both schemes $A$ and $B$ respectively. The interrelation is
given by the almost obvious relations
\begin{mathletters}
\label{X1}
\begin{equation}
x_B=T(u_{E}^{B})T^{-1}(u_{E}^{A})x_A,
\end{equation}
\begin{equation}
u_{E}^{B}=\Lambda_{BA}u^{A}_{E},
\end{equation}
\end{mathletters}
where $u^{A}_{E}(u^{B}_{E})$ is the four-velocity of $\Sigma_A(\Sigma_B)$
with respect to $\Sigma$ expressed in the EP synchronization for convenience.
$T(u_E)$ is given by the eq.\ (\ref{A18a}) (with $\vec{\sigma}_E$
replaced by the corresponding four-velocity $u_E$).  We observe that a set of
all possible four-velocities $u_E$ is related by Lorentz group
transformations too, i.e.\ $\{\Lambda_{BA}\}=L_S$.  Of course it does not
coincide with our intersystemic Lorentz group $L$.  We call the group $L_S$
a synchronization group \cite{Rem2}.

Now, if we compose the transformations (\ref{A1}, \ref{A2}) of $L$ and
(\ref{X1}) of $L_S$ we obtain
\begin{equation}
(\Lambda_S,\Lambda):\qquad
\left.\begin{array}{ll}
x'=T(\Lambda_S\Lambda u_E)\Lambda T^{-1}(u_E)x, \\
u_{E}'=\Lambda_{S}\Lambda u_E
\end{array}\right\} \label{X2}
\end{equation}
with $\Lambda_S\in L_S, \Lambda\in L$.

Thus the composition law for $(\Lambda_S,\Lambda)$ reads
\begin{equation}
(\Lambda_S',\Lambda')(\Lambda_S,\Lambda)=
(\Lambda_S'(\Lambda'\Lambda_S{\Lambda'}^{-1}),\Lambda'\Lambda). \label{X3}
\end{equation}
Therefore, in a natural way, we can select three subgroups:
\[
L=\{(I,\Lambda)\},\quad L_S=\{(\Lambda_S,I)\},\quad
 L_0=\{(\Lambda_{S},\Lambda^{-1}_{S})\}.
\]
By means of (\ref{X3}) it is easy to check that $L_0$ and $L_S$ commute.
Therefore the set $\{(\Lambda_S,\Lambda)\}$ is simply the direct product of
two Lorentz groups $L_0\otimes L_S$.  The intersystemic Lorentz group $L$ is
the diagonal subgroup in this direct product.  From the composition law
(\ref{X3}) it follows that $L$ acts as an authomorphism group of $L_S$.

Now, the synchronization group $L_S$ realizes in fact the relativity
principle.  In our language the relativity principle  can be formulated as
follows: {\em Any inertial frame can be choosen as a preferred frame}.
What happens, however, when the tachyons do exist? In that case the
relativity principle is obviously broken: {\em If tachyons exist then one
and only one inertial frame \underline{must be} a prefered frame}.  Moreover,
the one-way light velocity becomes a real, measured physical quantity
because conventionality thesis breaks down.  It means that the
synchronization group $L_S$ is broken to the $SO(3)_{u}$ subgroup
(stability group of $u_E$); indeed, transformations from the $L_S/SO(3)_{u}$
do not leave the causality notion invariant.  As we show later, on the
quantum level we have to deal with {\em spontaneous breaking\/} of $L_S$
to $SO(3)$.

\section{Quantization}\label{quant}
As was mentioned in Section \ref{CTsyn}, the following two facts, true only
in the CT synchronization, are extremaly important for a proper quantization
procedure; namely the invariance of the sign $\varepsilon(k^0)$ of $k^0$ and
the existence of a covariant lower energy bound.  The first one allows an
invariant decomposition of the tachyonic field into a creation and an
anihilation parts.  The second one is necessary to guarantee stability of the
quantum theory.  Recall that the noninvariance of $\varepsilon(k^{0}_{E})$
and the absence of a lower bound for $k_{0E}$ were the main reason why the
construction of a quantum theory for tachyons in the EP synchronization
scheme \cite{KK,KK2,Nak} was impossible.

This section is related to the approach presented in \cite{Rem2}.

\subsection{Local tachyonic field and its plane-wave decomposition}
\label{loc}
Let us consider a hermitean, scalar field $\varphi(x,\vec{\sigma})$
satysfying the corresponding Klein--Gordon equation with imaginary ``mass''
$i\kappa$, i.e.\
\begin{equation}
\left(g^{\mu\nu}(\vec{\sigma})\partial_{\mu}\partial_{\nu} -
\kappa^2\right) \varphi(x,\vec{\sigma})=0. \label{Q1}
\end{equation}
Our field $\varphi$ is $\vec{\sigma}$-dependent because (\ref{Q1}) is
assumed to be valid for an observer in an inertial frame moving with
respect to the privileged frame. Now, as in the standard case, let us
consider the Lorentz-invariant measure
\begin{equation}
d \mu(k,\vec{\sigma})= \theta(k^0) \delta(k^2+\kappa^2) d^4 k.
\label{Q2}
\end{equation}
Notice, that $d\mu$ does not have an analog in the EP synchronization
because of noninvariance of the sign of $k_{E}^{0}$ in this case.

The Heaviside step function $\theta(k^0)$ guarantees the positivity of $k^0$
and the lower bound of energy $k_0$ while $\delta(k^2+\kappa^2)$ projects on
the $\kappa^2$-eigenspace of the d'Alembertian
$g^{\mu\nu}\partial_{\mu}\partial_{\nu}$.  For this reason we can expand
invariantly the field $\varphi$ into the positive and negative frequencies
with respect to $k^0$
\begin{equation}
\varphi(x,\vec{\sigma})=\frac{1}{(2\pi)^{3/2}}\int d\mu(k,\vec{\sigma})\left(
e^{ikx}\,a^{\dagger}(k,\vec{\sigma})+e^{-ikx}\,a(k,\vec{\sigma})\right).
\label{22}
\end{equation}
Integrating with
respect to $k_0$ we obtain
\begin{equation}
\varphi(x,\vec{\sigma})=\frac{1}{(2\pi)^{3/2}}
\int_{\Gamma}\frac{d^3\nad{k}}{2\omega_k}
\left(e^{ik_{+}x}\,a^{\dagger}(k_{+},\vec{\sigma})+
e^{-ik_{+}x}\,a(k_{+},\vec{\sigma})\right).
\label{23}
\end{equation}
Here $k_{+}x=k_{0+}x^0+\nad{k}\vec{x}$ and $k_{0+}$ is a solution of the
dispersion relation $k^2=-\kappa^2$; namely $k_{0+}$, given in terms of
velocities by the eq.\ (\ref{D26}), has the following form with respect to
$\nad{k}$
\begin{equation}
k_{0+}=-\frac{\vec{\sigma}}{c}\nad{k}+\gamma_{0}^{2}\omega_{k}
\label{XX1}
\end{equation}
with
\begin{equation}
\omega_k=\gamma_{0}^{-2}\sqrt{\left(\frac{\vec{\sigma}\nad{k}}{c}\right)^2+
\left(|\nad{k}|^2-\kappa^2\right)\gamma_{0}^{2}}.
\label{XX2}
\end{equation}
Note that $k_{+}^{0}=\omega_{k}$.
The integration range $\Gamma$ is determined by
the constraint $k^2=-\kappa^2$, namely
\begin{equation}
|\nad{k}|\geq\kappa\left(1+\left(\gamma_{0}^{2}-1\right)
\left(\frac{\vec{\sigma}\nad{k}}
{|\vec{\sigma}||\nad{k}|}\right)^2\right)^{-1/2},
\label{Q3}
\end{equation}
i.e.\ values of $\nad{k}$ lie outside the oblate spheroid with halfaxes
$a=\kappa$ and $b=\kappa\gamma^{-1}_{0}$.  Note that $\Gamma$ is invariant
under the inversion $\nad{k}\rightarrow-\nad{k}$.

For the operators $a$ and $a^{\dagger}$ we postulate the canonical
commutation rules
\begin{mathletters} \label{25}
\begin{equation}
\left[a(k_{+},\vec{\sigma}),a(p_{+},\vec{\sigma})\right]
=\left[a^{\dagger}(k_{+},\vec{\sigma}),a^{\dagger}(p_{+},\vec{\sigma})\right]
= 0,
\end{equation}
\begin{equation}
\left[a(k_{+},\vec{\sigma}),a^{\dagger}(p_{+},\vec{\sigma})\right]
=2\omega_{k}\delta(\nad{k}-\nad{p}).
\end{equation}
\end{mathletters}
The vacuum $\left | 0 \right >$ is assumed to satisfy the conditions
\begin{equation}
\left < 0 | 0 \right > = 1 \quad \text{and}\quad
a(k_{+}, \vec{\sigma}) \left |0 \right >
= 0.  \label{26}
\end{equation}
By the standard procedure, using eq.\ (\ref{23}), we obtain
the commutation rule
for $\varphi(x, \vec{\sigma})$ valid for an arbitrary separation
\begin{equation}
\left [ \varphi(x, \vec{\sigma}), \varphi(y, \vec{\sigma}) \right ] = - i
\Delta(x - y, \vec{\sigma}),  \label{27}
\end{equation}
where the analogon of the Schwinger function reads
\begin{equation}
\Delta(x, \vec{\sigma}) = \frac{- i}{(2\pi)^{3}} \int d^{4}k\,
\delta(k^{2} + \kappa^{2})\, \varepsilon(k^{0})\, e^{ikx}.  \label{28}
\end{equation}
It is remarkable that $\Delta$ does not vanish for a space-like separation
which is a direct consequence of the faster-than-light propagation of the
tachyonic quanta. Moreover $\Delta(x, \vec{\sigma})|_{x^{0}=0} = 0$
and therefore no interference occurs between two measurements of $\varphi$
at an instant time. This property is consistent with our interpretation of
instant-time hyperplanes as the initial ones.

Now, because of the absolute meaning of the arrow of time in the CT
synchronization we can introduce an invariant notion of the time-ordered
product of field operators. In particular the tachyonic propagator
\[
{\Delta}_{F}(x-y,\vec{\sigma})=-i\left<0\right|
T(\varphi(x,\vec{\sigma})\,\varphi(y,\vec{\sigma}))\left|0\right>
\]
is given by
\begin{equation}
{\Delta}_{F}(x,\vec{\sigma})=-\theta(x^0)\,
{\Delta}^{-}(x,\vec{\sigma})+\theta(-x^0)\,{\Delta}^{+}(x,\vec{\sigma})
\label{29}
\end{equation}
with
\begin{equation}
{\Delta}^{\pm}(x,\vec{\sigma})=\frac{\mp i}{(2\pi)^3}
\int d^4k\,\theta(\pm k^0)\delta(k^2+\kappa^2)e^{ikx}.
\label{Q4}
\end{equation}
The above singular functions are
well defined as distributions on the space of ``well behaved'' solutions of
the Klein--Gordon equation (\ref{Q1}).

The role of the Dirac delta plays the generalized function
\begin{equation}
{\delta}^{4}_{\Gamma}(x-y)=\frac{1}{(2\pi)^3}\delta(x^0-y^0)
\int_{\Gamma}d^3\nad{k}\,e^{i\nad{k}(\vec{x}-\vec{y})}.  \label{30}
\end{equation}
The above form of ${\delta}^{4}_{\Gamma}(x)$ express  impossibility of the
localization of tachyonic quanta. In fact, the tachyonic field does not
contain modes with momentum $\nad{k}$ inside the spheroid defined
in eq.\ (\ref{Q3}). Consequently, by the Heisenberg uncertainty
relation, an exact localization of tachyons is impossible.

Note also that
\[
\partial^0\Delta(x-y,\vec{\sigma})\delta(x^0-y^0)={\delta}^{4}_{\Gamma}(x-y)
\]
so the equal-time canonical commutation relations for
 $\varphi(x,\vec{\sigma})$ and its conjugate momentum $\pi(x,\vec{\sigma})=
\partial^0\varphi(x,\vec{\sigma})$ have the correct form
\begin{mathletters} \label{q25}
\begin{equation}
\delta(x^0-y^0)\left[\varphi(x,\vec{\sigma}),\varphi(y,\vec{\sigma})\right]=
\delta(x^0-y^0)\left[\pi(x,\vec{\sigma}),\pi(y,\vec{\sigma})\right]=0,
\end{equation}
\begin{equation}
\left[\varphi(x,\vec{\sigma}),\pi(y,\vec{\sigma})\right]\delta(x^0-y^0)=
i{\delta}^{4}_{\Gamma}(x-y)
\end{equation}
\end{mathletters}
as the operator eqations in the space of states.

To do the above quantization procedure mathematically more
 precise, we can use wave
packets rather than the plane waves. Indeed, with a help of the measure
(\ref{Q2}) we can define the Hilbert space $H^{+}_{\vec{\sigma}}$ of
one particle states with the scalar product
\begin{equation}
(f, g)_{\vec{\sigma}} = \int d\mu(k, \vec{\sigma})\,
f^{\ast}(k,\vec{\sigma})\,
g(k,\vec{\sigma})<\infty.  \label{C2}
\end{equation}
Now, using standard properties of the Dirac delta we deduce
\begin{equation}
(f,g)_{\vec{\sigma}}=\int_{\Gamma}\frac{d^3\nad{k}}{2\omega_k}
f^{*}(k_+,\vec{\sigma}) g(k_+, \vec{\sigma}).
\label{XX3}
\end{equation}
It is remarkable that for $\xi\rightarrow\infty$, $\omega_k\rightarrow 0$,
so to preserve inequality $\|f\|^{2}_{\vec{\sigma}}<\infty$,  the wave
packets $f(k_+,\vec{\sigma})$ rapidly decrease to zero with
$\xi\rightarrow\infty$. This means physically that probability of
``momentum localization'' of a tachyon in the infinite velocity limit
is going to zero in agreement with our intuition.
As usually we introduce the smeared operators
\begin{equation}
a(f, \vec{\sigma}) = (2\pi)^{-3/2}\int d\mu(k,\vec{\sigma})\,
a(k,\vec{\sigma})f^{\ast}(k, \vec{\sigma})  \label{C3}
\end{equation}
and the conjugate ones. The canonical commutation rules (\ref{25})
take the form
\begin{mathletters}\label{C4}
\begin{equation}
\left[a(f, \vec{\sigma}), a(g, \vec{\sigma})\right] =
\left[a^{\dagger}(f, \vec{\sigma}), a^{\dagger}(g,\vec{\sigma})\right] = 0,
\end{equation}
\begin{equation}
\left[a(f, \vec{\sigma}), a^{\dagger}(g, \vec{\sigma})\right] =
(f, g)_{\vec{\sigma}}.
\end{equation}
\end{mathletters}
We have also $a(f,\vec{\sigma})\left|0\right> = 0$ and
$\left<f,\vec{\sigma}|g,\vec{\sigma}\right>=\left(f,g\right)_{\vec{\sigma}}$,
where
$\left|f,\vec{\sigma}\right>=a^{\dagger}(f,\vec{\sigma})\left|0\right>$.
Let us discuss the implementation of the intersystemic Lorentz group
$L$ on the quantum level.  According to our assumption of scalarity of
$\varphi(x,\vec{\sigma})$
\begin{equation}
L\ni\Lambda:\;
\varphi'(x',\vec{\sigma}')=\varphi(x,\vec{\sigma}).  \label{A}
\end{equation}
where $x'$ and $\vec{\sigma}'$ are given by (\ref{D15}--\ref{D16}).
The transformation law should be realized by a reprezentation
$U(L)$ as follows
\begin{equation}
U(\Lambda)\varphi(x,\vec{\sigma})U^{-1}(\Lambda)=\varphi(x',\vec{\sigma}'),
\label{C}
\end{equation}
i.e.\
\begin{equation}
U(\Lambda)a(k,\vec{\sigma})U^{-1}(\Lambda)=a(k',\vec{\sigma}')
\label{D}
\end{equation}
and
\begin{equation}
U(\Lambda)\left|0\right>=\left|0\right>.
\label{E}
\end{equation}
Therefore the wave packets must satisfy the scalarity condition
(\ref{A}) i.e.\
\begin{equation}
f'(k',\vec{\sigma}')=f(k,\vec{\sigma}).
\label{F}
\end{equation}
It follows that the family $\{U(\Lambda)\}$ forms an unitary representation
of $L$; indeed we see that
\begin{equation}
\left(f',g'\right)_{\vec{\sigma}'}=\left(f,g\right)_{\vec{\sigma}}.
\label{H}
\end{equation}
Summarizing, the Lorentz group $L$ is realized by a family of unitary
mappings in the following bundle of Hilbert spaces
\begin{itemize}
\item[] $H_0$ (vacuum);
\item[] $H^{+}_{\vec{\sigma}}$ (bundle of one-particle spaces of states);
\item[] $H^{+}_{\vec{\sigma}}\otimes H^{+}_{\vec{\sigma}}$ (bundle of
two-particle spaces of states);
\item[] \dotfill\ etc.\ \hspace*{\fill}
\end{itemize}
i.e.\ $H^{+}=H_0\oplus\bigcup_{\vec{\sigma}}H^{+}_{\vec{\sigma}}
\oplus\bigcup_{\vec{\sigma}}H^{+}_{\vec{\sigma}}
\otimes H^{+}_{\vec{\sigma}}
\oplus\dots$ etc.\ with the base space as the velocity space
($\vec{\sigma}$-space).  Now we introduce wave-packet
solutions of the Klein--Gordon equation {\em via\/} the Fourier
transformation
\begin{equation}
{\cal F}(x,\vec{\sigma}) = (2 \pi)^{-3/2} \int d\mu(k,\vec{\sigma})\,
f(k,\vec{\sigma})e^{-ikx}.  \label{C5}
\end{equation}
In terms of these solutions the scalar product (\ref{C2}) reads
\begin{equation}
({\cal F},{\cal G})_{\vec{\sigma}} = - i \int d^{3}\vec{x}\,
{\cal F}^{\ast}(x,\vec{\sigma})\stackrel{\leftrightarrow}{\partial}^0
{\cal G}(x,\vec{\sigma}).  \label{C6}
\end{equation}
It is easy to see that for an orthonormal basis
$\{\Phi_{\alpha}(x, \vec{\sigma})\}$ in
$H^{+}_{\vec{\sigma}}$  the
completeness relation holds
\begin{equation}
\sum_{\alpha} \Phi^{\ast}_{\alpha}(x, \vec{\sigma})
\Phi_{\alpha}(y, \vec{\sigma}) = i \Delta^{+}_{T}(x - y, \vec{\sigma}),
 \label{C7}
\end{equation}
where $\Delta^{+}$ has the form (\ref{Q4}) and it is the reproducing
kernel in $H^{+}_{\vec{\sigma}}$ i.e.\
\[
(i \Delta^{+}(x), \Phi)_{\vec{\sigma}} = \Phi(x, \vec{\sigma}).
\]
Finally, translational invariance implies the following, almost
standard, form of the fourmomentum operator
\begin{equation}
P_{\mu}=\int d\mu(k,\vec{\sigma})\,k_{\mu} a^{\dagger}(k,\vec{\sigma})
a(k,\vec{\sigma}).\label{XXX1}
\end{equation}
It is evident that the vacuum $\left| 0 \right>$ has zero fourmomentum.
Furthermore, $P_{\mu}$ applied to one-particle state
$a^{\dagger}(k_+,\vec{\sigma})\left| 0
\right>=\left|k_+,\vec{\sigma}\right>$ gives
\begin{equation}
P_{\mu} \left|k_+,\vec{\sigma}\right>=k_{\mu+}\left|k_+,\vec{\sigma}\right>
\label{XX4}
\end{equation}
Recall that the asymptotic  one-particle lower energy bound
\[
\lim_{\xi\rightarrow\infty}k_{0+}=\frac{s-1}{\sqrt{2s-1}},
\]
although can be negative, it is always finite because
\[
\frac{-\varepsilon}{\sqrt{1-\varepsilon^2}}<\frac{s-1}{\sqrt{2s-1}}<
\frac{\varepsilon}{1-\varepsilon^2}
\]
where $\varepsilon=\frac{\sigma_E}{c}\in\left<0,1\right)$ is fixed in
a fixed inertial frame. Moreover, the average over all directions
of the minimal asymptotic energy
is equal exactly to zero.
Thus we have constructed a consistent quantum field theory for the
hermitean, scalar tachyon field $\varphi(x,\vec{\sigma})$. We
conclude, that a proper framework to do this is the CT
synchronization scheme.

\subsection{Spontaneous breaking of the synchronization group}\label{break}
As we have seen in the foregoing section, intersystemic Lorentz group
is realized unitarily on the quantum level. In this section we will
analyze possibility of incorporation of the synchronization group $L_S$ in
our scheme.

As was stressed in the Sec.\ \ref{syn}, if tachyons exist then
one and only one inertial frame is the preferred frame. In other words the
relativity principle is broken in this case: tachyons distinguish a fixed
synchronization scheme from the family of possible CT synchronizations.
Consequently, because all admissible synchronizations are
related by the group $L_S$, it should be broken. To see
this let us consider transformations belonging to the $L_0$ subgroup
(see Sec.\ \ref{syn}). They are composed from the transformations of
intersystemic Lorentz group $L$ and the synchronization group $L_S$;
namely they have the following form (see eq.\ (\ref{X2}) and the
definition of $L_0$),
\begin{equation}
\vec{\sigma}'=\vec{\sigma},\qquad x'=T(\vec{\sigma})\Lambda^{-1}_{S} T^{-1}
(\vec{\sigma})x\equiv\Lambda^{-1}_{S}(\vec{\sigma})x. \label{ZZ}
\end{equation}
We search an operator $W(\Lambda)$ implementing (\ref{ZZ}) on the
quantum level; namely
\begin{equation}
\varphi'(x,\vec{\sigma})=W(\Lambda_{S})\varphi(x,\vec{\sigma})W^{\dagger}
(\Lambda_S)=\varphi(x',\vec{\sigma}) \label{ZZZ}
\end{equation}
This means that we should compare both sides of (\ref{ZZZ}) i.e.
\[
\int d\mu(k,\vec{\sigma})\left[ e^{ikx}a^{\prime\dagger}
(k,\vec{\sigma})+e^{-ikx}a^{\prime\dagger}(k,\vec{\sigma})\right]
\]
\begin{equation}
=\int d\mu(p,\vec{\sigma})\left[
e^{ipx'}a^{\dagger}
(p,\vec{\sigma})+e^{-ipx'}a^{\dagger}(p,\vec{\sigma})\right],
\label{84}
\end{equation}
where $x'$ is given by eq.\ (\ref{ZZ}), while, formally
\begin{equation}
a'=WaW^{\dagger},\qquad a^{\prime\dagger}=Wa^{\dagger}W^{\dagger}.
\label{85}
\end{equation}
Taking into account the form of the measure $d\mu$ (eq.\ (\ref{Q2}))
and the fact that $\Lambda_{S}(\vec{\sigma})$ does not leave invariant
the sign of $k^0$, after some calculations, we deduce the following
form of $W$:
\begin{mathletters}\label{86}
\begin{equation}
a'(k,\vec{\sigma})=\theta(k^{\prime 0})a(k',\vec{\sigma})+
\theta(-k^{\prime 0})a^{\dagger}(-k',\vec{\sigma}),
\end{equation}
\begin{equation}
a^{\prime\dagger}(k,\vec{\sigma})=\theta(k^{\prime 0})a^{\dagger}
(k',\vec{\sigma})+
\theta(-k^{\prime 0})a(-k',\vec{\sigma}),
\end{equation}
\end{mathletters}
where $k'=\Lambda_S(\vec{\sigma})k$.

We see that formally unitary operator $W(\Lambda_S)$ is realized by
the Bogolubov-like transformations; the Heaviside $\theta$-step
functions are the Bogolubov coefficients. The form (\ref{86}) of the
transformations of the group $L_0$ reflects the fact, that
a possible change of the
sign of $k^0$ causes a different decomposition
of the field $\phi$ on the positive and negative frequencies.
Furthermore it is easy to check that the transformation (\ref{86}) preserves
the canonical commutation relations (\ref{25}).

However, the formal operator $W(\Lambda_S)$ realized in the ring of
the field operators, cannot be unitarily implemented in the space of
states in general; only if $\Lambda=\Lambda_u$ is an
element of the stability group $SO(3)_u$ of $u$ in $L_S$, it can be
realized unitarily. This is related to the fact that
$\Lambda_u(\vec{\sigma})$ does not change the sign of $k^0$ for any
$k$. Indeed, notice firstly that for $\Lambda_S\in L_S/SO(3)$,
$W(\Lambda_S)$ does not anihilate the vacuum $\left|0\right>$.
Moreover, the particle number operator
\begin{equation}
N=\int d\mu(k,\vec{\sigma}) a^{\dagger}(k,\vec{\sigma}) a(k,\vec{\sigma})
\label{87}
\end{equation}
applied to the ``new'' vacuum
\begin{equation}
\left|0\right>'=W^{-1}\left|0\right>   \label{88}
\end{equation}
gives
\begin{equation}
N\left|0\right>'=\delta^3(0)\int_{\Gamma}
d^3\nad{k}\,\theta(-(\Lambda_S(\vec{\sigma})k_+)^0)\left|0\right>'.
\label{89}
\end{equation}
The right side of the above expression diverges like $\delta^6(0)$ for
any $\Lambda_S(\vec{\sigma})\in L_S/SO(3)_u$. Only for the stability
subgroup $SO(3)_u\subset L_S$ vacuum remains invariant. Thus, a
``new'' vacuum $\left|0\right>'$, related to an esssentialy new
synchronization, contains an infinite number of ``old'' particles. As
is well known, in such a case, two Fock spaces $H$ and $H'$, generated
by creation operators from $\left|0\right>$ and $\left|0\right>'$
respectively, cannot be related by an unitary
transformation\footnote{We can treat (\ref{86}), in some sense, as a
quantum version of the familiar {\em reinterpretation
principle\/} \cite{Fei}. We find that the reinterpretation principle cannot
be unitarily implemented.} ($W(\Lambda_S)$ in our case). Therefore, we
have deal with the so called spontaneous symmetry breaking of $L_S$
to the stability subgroup $SO(3)$. This means that
physically privileged is only one realization of the cannonical
commutation relations (\ref{25}) corresponding to a vacuum
$\left|0\right>$ defined by eq.\ (\ref{26}). Such a realization is related
to a definite choice of the privileged inertial frame and
consequently to a definite CT synchronization scheme. Thus we can
conclude that {\em tachyons distinguish a preferred frame via
spontaneous breaking of the synchrony group}.

To complete discussion, let us apply the fourmomentum operator $P_{\mu}$ to
the new vacuum $\left|0\right>'$. As the result we obtain
\begin{equation}
P_{\mu}\left|0\right>'=-\delta^3(0){{\Lambda_S}_\mu}^{\nu}(\vec{\sigma})
\int_{\Gamma}d^3\nad{k}\,\theta(-(\Lambda_S(\vec{\sigma})k_+)^0)k_{\nu}
\left|0\right>'.  \label{90}
\end{equation}
This expression diverges again like $\delta^7(0)$ for $\Lambda_S\in
L_S/SO(3)_u$. Therefore a transition to a new vacuum ($\equiv$
change of the privileged frame) demands an infinite momentum transfer,
i.e.\ it is physically unadmissible. This last phenomenon supports our
claim that existence of tachyons is associated with spontaneous breaking of
the the synchronization group.

On the other hand it can be simply shown \cite{Rem2} that a free field theory
for standard particles (bradyons or luxons), formulated in CT
synchronization, is unitarily equivalent to the standard field theory in the
EP synchronization.

\section{Conclusions}\label{concl}
We can conclude that, contrary to the current opinion, a consistent
quantization of the tachyonic field in the framework of special relativity is
possible and it is closely connected with the choice of an appriopriate
synchronization scheme.  From this point of view the Einstein--Poincar\'e
synchronization is useless in the tachyonic case.  On the other hand, in a
description of bradyons and luxons only, we are free in the choice of a
synchronization scheme.  For this reason we can use in this case
CT-synchronization as well as the standard one.

The CT-synchronization, a natural one for a description of tachyons,
favourizes a reference frame (privileged frame).  This preference is only
formal if tachyons do not exist.  However, if they exist, then an inertial
reference frame is really (physically) preferred.  As a consequence, the
one-way light velocity can be measured in this case and, in general, it will
be direction-dependent for a moving observer.  Light velocity is isotropic
only in the privileged frame.

A next step is to construct a field theory for a fermionic tachyon and a
local interaction with other fields.

Two questions arise immediately.  Do tachyons exist?  Are candidates from the
family of observable particles to be tachyons?

An answer to the above questions is not so simple. For example, about the
existence of tachyons we can deduce indirectly. Indeed, we know that they
prefer an inertial frame (relativistic ether). On the other hand we have in
the real word a serious candidate to such a frame; namely frame related
to the background radiation. Moreover, the standard cosmological
model and related models possess in fact an absolute time (radius of the
universe). Furthermore, let us notice that if the one-way velocity is really
anisotropic, then our present interpretation of the experimental data
about of distribution of cosmic matter, its anisotropy and anisotropy of
the background radiation, is in some sense false and demands a
reinterpretation.  Indeed, our actual picture assumes isotropic one-way light
velocity equal to $c$. Maybe, by means of anisotropy of
the one-way light velocity it is possible to obtain a more isotropic
picture of the world.

Of course, indirect arguments are not decisive ones.  An experimental
evidence for tachyons can be a decisive argument only.  However, up to now,
in the current opinion, no candidates to be a tachyon.  Notwithstanding,
there is a number of experiments \cite{tab,Ass} suggesting that the muonic
neutrino mass square is negative by a few standard deviations.
The experimental data of the electron neutrino mass square,
although not so dramatic, also prefers a negative value \cite{tab}.
Maybe some decisive data will be obtained after
more accurate measurements of the
charged pion mass which have began now at the Paul Scherrer Institute
\cite{tab,Ass}.
If the future results will support intriguing hypothesis by Chodos {\em et
al.} that neutrino is a fermionic tachyon, then a modification of the theory
of electroweak interactions will be necessary.


\begin{thebibliography}{10}

\bibitem{Rec1}
E. Recami, Giornale di Fisica {\bf 10},  195  (1968).

\bibitem{OR}
V. Olkhovsky and E. Recami, Nuov.\ Cim. {\bf A63},  814  (1969).

\bibitem{Rec2}
E. Recami, Rev.\ del Nuov.\ Cim. {\bf 9},  1  (1986).

\bibitem{KK}
K. Kamoi and S. Kamefuchi, Prog.\ Theor.\ Phys. {\bf 45},  1646  (1971).

\bibitem{KK2}
K. Kamoi and S. Kamefuchi,  in {\em Tachyons, Monopoles and Related Topics},
  edited by E. Recami (North-Holland, New York, 1978).

\bibitem{tab}
{Particle Data Group}, Phys. Rev. {\bf D50},  1390  (1994).

\bibitem{Ass}
K. {Assamagan {\em et al.}}, Phys. Lett. {\bf B335},  231  (1994).

\bibitem{CHK}
A. Chodos, A.~I. Hauser, and V.~A. Kostelecky, Phys. Lett. {\bf B150},  431
  (1985).

\bibitem{Bie}
L.~C. Biedenharn, H. van Dam, and Y.~J. Ng, Phys. Lett. {\bf B158},  227
  (1985).

\bibitem{Bar}
A. Barut,  in {\em Tachyons, Monopoles and Related Topics}, edited by E. Recami
  (North-Holland, New York, 1978).

\bibitem{KS}
V.~A. Kostelecky and S. Samuel, Phys. Rev. {\bf D42},  1289  (1990).

\bibitem{KP}
V.~A. Kostelecky and M.~J. Perry, Nucl.\ Phys. {\bf B414},  174  (1994).

\bibitem{CG}
C.~G. Callan and Z. Gan, Nucl.\ Phys. {\bf B272},  647  (1986).

\bibitem{DS}
S.~R. Das and B. Sathiapalan, Phys. Rev. Lett. {\bf 56},  2664  (1986).

\bibitem{Tse}
A.~A. Tseytlin, Phys. Lett. {\bf B264},  311  (1991).

\bibitem{MM}
S. Mahapatra and S. Mukherji, Preprint IC/94/116, ICTP, Trieste (unpublished).

\bibitem{Tan}
F.~R. Tangherlini, Nuov.\ Cim. Suppl. {\bf 20},  1  (1961).

\bibitem{Cha1}
T. Chang, Phys. Lett. {\bf A70},  1  (1979).

\bibitem{Cha2}
T. Chang, J. Phys. {\bf A13},  L207  (1980).

\bibitem{Rem1}
J. Rembieli{\'n}ski, Phys. Lett. {\bf A78},  33  (1980).

\bibitem{Fei}
G. Feinberg, Phys. Rev. {\bf 159},  1089  (1967).

\bibitem{Rem2}
J. Rembieli{\'n}ski, Preprint IF U{\L}/1/1980, Instytut Fizyki Uniwersytetu
  {\L}{\'o}dzkiego (unpublished).

\bibitem{Ter}
Y.~P. Terletsky, Sov.\ Phys.\ Dokl. {\bf 5},  782  (1960).

\bibitem{Tnk}
S. Tanaka, Prog.\ Theor.\ Phys. {\bf 24},  171  (1960).

\bibitem{BDS}
O.~M.~P. Bilaniuk, V.~K. Deshpande, and E.~C.~G. Sudarshan, Am.\ Journ.\ Phys.
  {\bf 30},  718  (1962).

\bibitem{T&M}
E. Recami,  in {\em Tachyons, Monopoles and Related Topics}, edited by E.
  Recami (North-Holland, New York, 1978).

\bibitem{Nak}
N. Nakanishi, Prog.\ Theor.\ Phys. Suppl. {\bf 51},  1  (1972).

\bibitem{Cha3}
T. Chang, J. Phys. {\bf A12},  L203  (1979).

\bibitem{Rei}
H. Reichenbach, {\em Axiomatization of the Theory of Relativity} (University of
  California Press, Berkeley, CA, 1969).

\bibitem{VT}
J.~G. Vargas and D.~G. Torr, Found.\ of Phys. {\bf 16},  1089  (1986).

\bibitem{Var}
J.~G. Vargas, Found.\ of Phys. {\bf 16},  1003  (1986).

\bibitem{Jam}
M. Jammer,  in {\em Problems in the Foundations of Physics} (North-Holland,
  Bologne, 1979).

\bibitem{Sjo}
T. Sj{\"o}din, Nuov.\ Cim. {\bf B51},  229  (1979).

\bibitem{And}
J.~L. Anderson, {\em Principles of Relativity Physics} (Academic Press, New
  York, 1967).

\bibitem{SP}
T. Sj{\"o}din and M.~F. Podlaha, Lett.\ al Nuov.\ Cim. {\bf 31},  433  (1981).

\bibitem{FN}
A. Flidrzy{\'n}ski and A. Nowicki, J. Phys. {\bf A15},  1051  (1982).

\end{thebibliography}

\end{document}